# Pellet fuelling of plasmas with ELM mitigation by resonant magnetic perturbations in MAST

**M Valovič, G Cunningham, L Garzotti, C Gurl, A Kirk, G Naylor, A Patel, R Scannell, A J Thornton and the MAST team**

EURATOM/CCFE Fusion Association, Culham Science Centre,
Abingdon, OX14 3DB, UK

email: martin.valovic@ccfe.ac.uk

**Abstract.** Shallow fuelling pellets are injected from the high field side into plasmas in which ELMs have been mitigated using external magnetic perturbation coils. The data are compared with ideal assumptions in the ITER fuelling model, namely that mitigated ELMs are not affected by fuelling pellets. Firstly it is shown that during the pellet evaporation an ELM is triggered, during which the amount particle loss could be larger (factor ~1.5) than the particle loss during an ELM which was not induced by pellet. Secondly, a favourable example is shown in which post-pellet particle losses due to mitigated ELMs are similar to the non-pellet case, however unfavourable counter-examples also exist.

## 1. Introduction

In future tokamak fusion devices like ITER the frequency of naturally occurring edge localised modes (ELMs) is expected to be low causing large modulations of the energy flux that can not be handled by the divertor. One of the techniques for active control of the ELM frequency which has been successfully demonstrated, and is considered for ITER, is the application of resonant magnetic perturbations (RMPs).

An important step in the development of ELM mitigation is to test its compatibility with density control as ELM mitigation often affects the particle confinement, e. g. see [1]. In ITER, the main density control tool is likely to be the injection of frozen deuterium pellets from the high field side (HFS) of the plasma. First data on pellet fuelling with ELM mitigation by RMPs have already been collected from major tokamaks. On JET low field side pellets have been used to refuel plasmas with RMPs resulting in a further increase of ELM frequency and an additional reduction of power deposited to the outer divertor targets during ELMs [2]. On DIII-D pellet fuelling of a plasma with fully suppressed ELMs sometimes results in a return to ELMy H-mode [3]. On ASDEX-Upgrade, with pellet fuelling into a discharge in which type I ELMs have been suppressed, some energy losses synchronous with pellets were observed [4]. Note that the pedestal collisionality is lower and pellet deposition is shallower in the DIII-D case compared to the ASDEX-Upgrade experiment.

This paper presents the results of the first experiments on the Mega Amp Spherical Tokamak (MAST) on the interaction of HFS pellets with RMP ELM mitigation. The focus is to examine the deviation from the pellet fuelling model which assumes that ELM mitigation is independent of fuelling pellets.

## 2. Pellet fuelling model

Before presenting the experimental results it is useful to define the ideal state of affairs against which the experimental data can be compared. This model for ITER has been described previously in dedicated papers [5, 6]. Here we present its simplified version in which we assume that ELMs are mitigated by resonant magnetic perturbations and not by pellet ELM pacing. This means that we are not concerned about competition between pellet fuelling and pellet ELM pacing for overall particle throughput. The starting point of the model is the assumption that we have perfect control over the ELM frequency $f_{ELM}$. This frequency is set to the value:

$$f_{ELM} = \alpha P / \delta W_{ELM} \tag{1}$$

where $P$ is the power loss, $\alpha$ is the fraction of power loss due to ELMs [7] and $\delta W_{ELM}$ is the maximum energy loss per ELM which can be handled by a divertor. For the present ITER divertor $\delta W_{ELM} = 0.6$ MJ so that for the standard scenario with $P = 100MW$ the ELM frequency is $f_{ELM} \sim 33$Hz (with $\alpha = 0.2$) [6]. The next step is to assume that the number of particles lost per ELM, $\delta N_{ELM}$, is proportional to the relative energy loss (so called convective ELMs):

$$\delta N_{ELM} = \frac{\delta W_{ELM}}{W_{ped}} N_{ped} = \frac{\delta W_{ELM}}{T_{ped}} \tag{2}$$

Here $N_{ped}$ and $W_{ped}$ are the particle and energy content of plasma related to the pedestal respectively and $T_{ped}$ is the pedestal temperature. The assumption (2) that ELMs are conductive represents the most unfavourable case from the fuelling point of view. However in present devices convective ELMs are observed mostly at high collisionality and as the collisionality decreases significant temperature drop is observed during ELMs suggesting that conductive loss mechanism is dominant [8, 9]. Therefore on ITER with low collisionality pedestal $\delta N_{ELM} / N_{ped} < \delta W_{ELM} / W_{ped}$ is expected. Nevertheless to asses the upper limit for pellet fuelling requirement for ITER one usually takes the case of convective ELMs [6] so that $\delta N_{ELM} / N_{ped} = \delta W_{ELM} / W_{ped} = 0.006$.

The final step of the model is to assume a steady state situation so that the particle loss by ELMs is balanced by the pellet fuelling rate $\Phi_{pellet}$:

$$\Phi_{pellet} = f_{pellet} N_{pellet} = f_{ELM} \delta N_{ELM} \tag{3}$$

Here $f_{pellet}$ is the frequency of fuelling pellets and $N_{pellet}$ is the pellet particle content. The pellet diameter in ITER is set to 5mm ($N_{pel} / N_{ped} \sim 0.03$) so that the pellet frequency is $f_{pellet} \sim 8$Hz [5, 6].

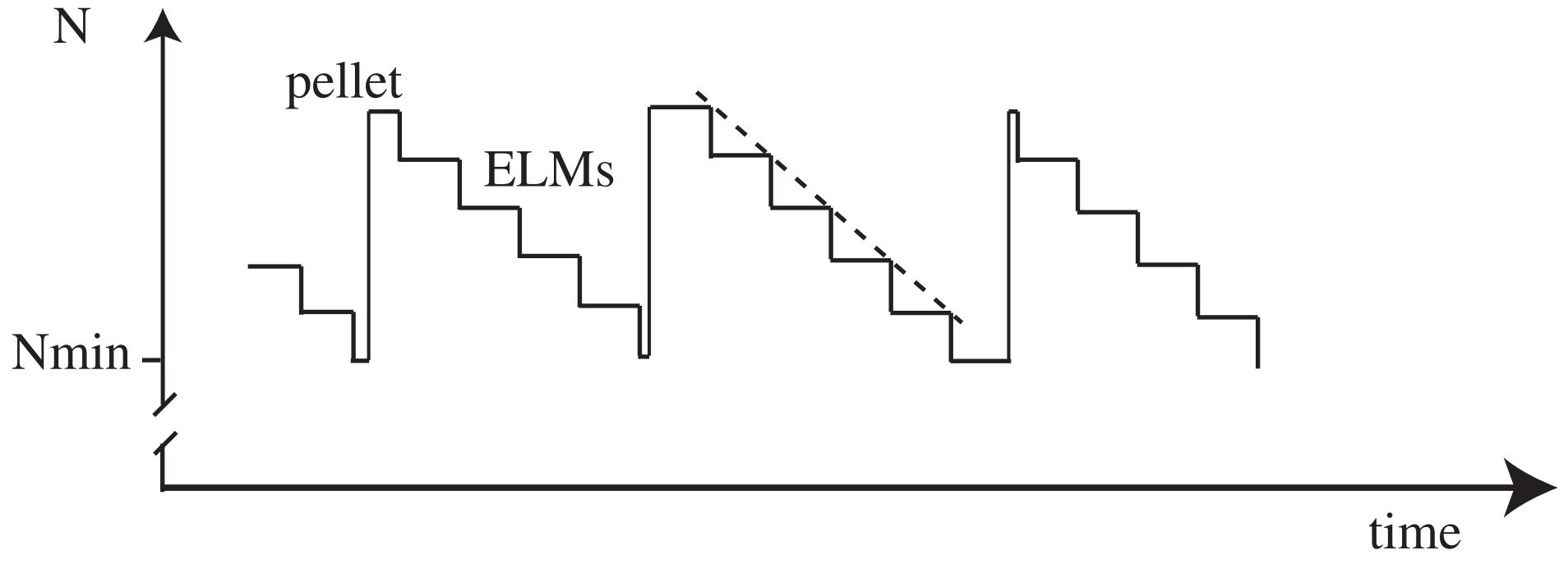

**Figure 1.** Schematics of the idealised pellet fuelling and ELM mitigation. Note that pellets and ELMs are not synchronised and the ELM-averaged particle loss is constant (indicated by broken line).

Figure 1 schematically illustrates the ideal pellet fuelling and ELM mitigation scenario. The density feedback system controls the density at $N = N_{\min}$. In this regime all ELMs are equal in size. This model implicitly assumes that pellets are small compared to the plasma particle content which means that the pedestal temperature is not modulated over the pellet cycle ($T_{ped} = \mathrm{const}(t)$) so that the energy loss per ELM is also constant ($\delta W_{ELM} = \mathrm{const}(t)$) as follows from equation (2).

The situation described above represents the ideal state of affairs: Particle and heat flux arrive to the divertor in small and equal (for each ELM) amounts. ELMs and pellets are not synchronised. When averaged over the ELM cycle, these fluxes are not modulated due to discrete pellet fuelling i.e. the post pellet density decay is linear (dashed line in figure 1).

The experiment can deviate from such ideal situation for many reasons. The most likely ones are:

- ELMs become synchronised with pellets
- Particle losses immediately after the pellet are enhanced

These effects can result in transients which counterbalance the ELM mitigation effort. In addition these deviations can reduce pellet fuelling efficiency and result in higher pellet fuelling rate, if density is controlled by feedback. Higher $\Phi_{pellet}$ will reduce pedestal temperature because of the relationship between particle and heat fluxes, $\Phi_{pellet} = \alpha P / T_{ped}$, if power is fixed. Lower pellet fuelling efficiency will also reduce the burn-up fraction of a fusion reactor.

## 3. Experimental conditions

The plasma used in this study has a single null divertor configuration (major radius $R_{geo} = 0.88\mathrm{m}$, minor radius $a = 0.49\mathrm{m}$, elongation $\kappa = 1.65$, plasma current $I_p = 550 - 640\mathrm{kA}$, toroidal magnetic field $B_T = 0.45\mathrm{T}$). The plasma is heated by neutral beams. ELMs are mitigated with RMP coils in an $n = 6$ configuration with coil current $I_{RMP} = 5.6$ kA-turns (4 turns), located at the lower-outer side of the plasma. This is the best plasma and RMP setup for ELM mitigation so far (for details see reference [10]). Pellets are injected from the top-high field side of the plasma (figure 2a). The pellet injector produces cylindrical pellets with nominal diameter and length $d_{pel} = L_{pel} = 1.3$ mm ($N_{pel} = 1.3 \times 10^{19}$ atoms), and velocities ~300m/s. For more details on pellet injection see [11]. Deuterium is used for gas fuelling, pellets and neutral beams.

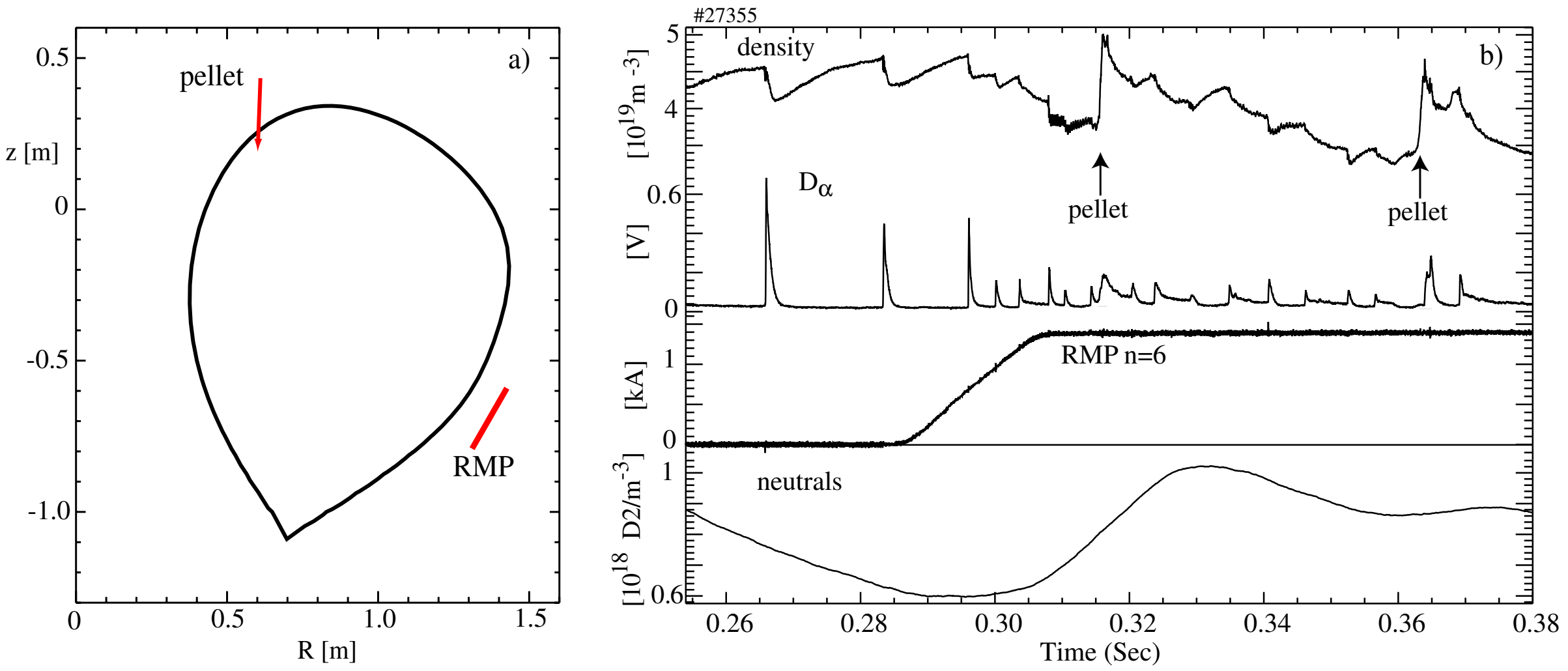

**Figure 2.** (a) Plasma cross section, geometry of pellet injection and RMP coils. (b) Typical waveforms for ELM mitigation and pellet fuelling experiment (from top to bottom): line averaged density, $D_\alpha$ emission, RMP current and neutral density in the vacuum vessel. Plasma current $I_p = 640\mathrm{kA}$ and injected neutral beam power $P_{NBI} = 1.9\mathrm{MW}$.

Figure 2b shows typical waveforms in the experiment. Application of RMPs increases the ELM frequency, decreases their amplitude and simultaneously decreases the plasma density. This “density pump-out” is partially compensated by an additional gas puff as seen from the increased main chamber neutral gas density. Pellets are injected during the flat top of the RMP current. Gas and pellet fuelling was used in “feed forward” mode based on experience from previous plasmas.

## 4. Mitigated ELMs without pellets

To assess the effect of pellets on ELMs, it is necessary to have a non-pellet comparison. Figure 3a shows the analysis of mitigated ELMs without pellet, using a fast interferometer signal. The ELM frequency is $f_{ELM} = 250\mathrm{Hz}$. The waveform of line integrated density $nL$ shows two very clear break-in-slope points during an ELM [12]. This suggests that the ELM particle losses occur during the well defined time interval. Good temporal localisation allows evaluation of the number of particles lost per ELM: $\delta(nL)_{ELM}/(nL)=3.2\%$ ($nL = 3.6\times10^{19}\,\mathrm{m}^{-2}$). The particle loss associated with ELMs without pellets is then:

$$\Phi_{ELM,nopel} \propto f_{ELM}\delta(nL)_{ELM} = 2.9\times10^{20}\,\mathrm{m}^{-2}\mathrm{s}^{-1} \qquad (4)$$

Figure 3b shows the Thomson scattering profiles taken just before and after the ELMs in figure 3a. Such measurement is enabled by operating this diagnostic in burst mode. It is seen that the radial extent of the ELM affected area is $\delta r_{ELM}/a \sim 0.18$.

In this paper we concentrate only on fuelling aspect of the ELM mitigation problem and by ELM size we mean the particle loss and not the energy loss. Plasma used as a target for pellet fuelling in this work is part of a larger dataset of plasmas with RMP ELM mitigation on MAST. This dataset is described in the specialised paper [13] which includes detailed analysis of energy loss per ELM and its parametric dependencies. Here we restrict ourselves only to showing the changes of electron temperature and electron pressure profiles during the ELMs (figures 3c and 3d). It is seen that the relative drop of electron density and electron temperature are comparable and both contribute to the energy loss.

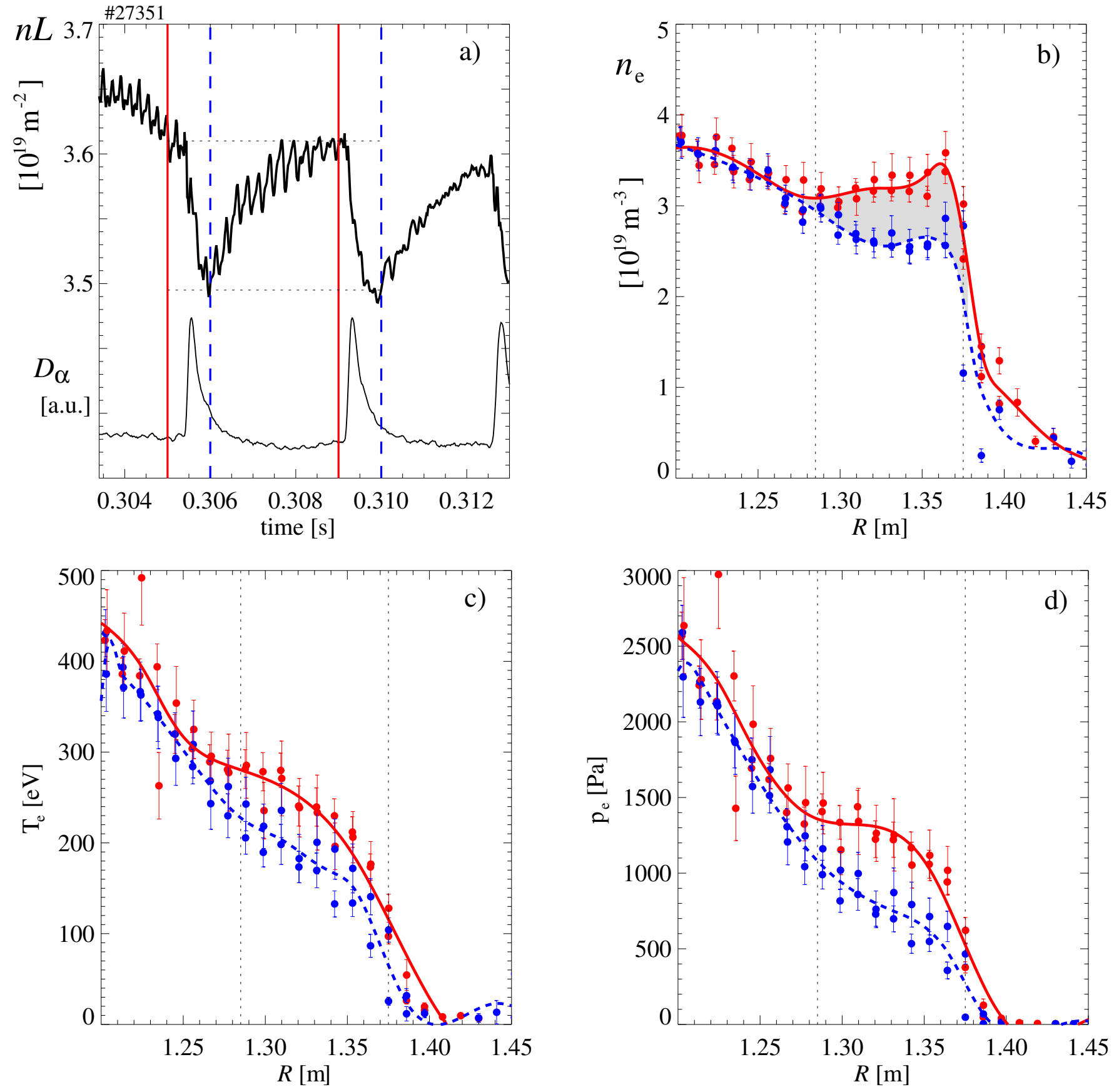

**Figure 3.** The size and affected area of ELMs mitigated by RMP. (a) line integrated density $nL$ from fast interferometer, $D_{\alpha}$ emission and vertical lines showing the timing of Thomson scattering. (b) electron density, (c) electron temperature and (d) electron pressure profiles from Thomson scattering before (red-solid line) and after (blue-dotted line) the ELMs. 2 ELMs are overlaid. $I_p = 640\text{kA}$, $P_{NBI} = 1.9\text{MW}$. Interferometer and Thomson scattering measurements are taken along the major radius with vertical offset of $\Delta z = 0.27\text{m}$ relative to the magnetic axis.

## 5. Pellet - ELM synchronisation

The first possible deviation from the pellet fuelling model presented in section 2 is the ELM triggering by pellets. Figure 4 shows the data from 5 pellets on a relative time scale. Figure 4a shows the increment of interferometer signal $\Delta(nL)$ during the pellet evaporation and deposition. This process lasts $\delta t_{pel} = 0.8 - 1.6\text{ms}$. Figure 4b shows the $D_{\alpha}$ emission from the lower outer leg of the divertor which is sensitive to ELMs but not to the light from the evaporating pellet. It is seen that for each pellet there is an ELM inside the time interval of pellet evaporation. A closer look shows that the ELMs are synchronised not with the beginning, but with the end of the pellet evaporation process.

Because the ELM frequency in ITER is about 5 times larger than the pellet frequency, the extra ELMs triggered by pellets increase the ELM frequency by 20% and thus should not cause a significant problem. This however assumes that the triggering of an ELM by the pellet is not increasing the ELM size. Direct measurement of the particle loss due to the pellet triggered ELM is complicated by the fact that the density profile during pellet evaporation is 3 dimensional due to the existence of an intense local particle source from the pellet. This is illustrated in figure 4c showing clear in-out asymmetry in the density profiles during the pellet deposition. The effect of the ELM is clearly seen from the difference between pre- and post ELM density profiles at the outer part of the plasma (shaded area in figure 4c). The change in the line integrated density due to the ELM at the outer part of the plasma is $\delta(nL)_{ELM,outer} = 8.3\times10^{17}\,\mathrm{m}^{-2}$. This is 1.5 times larger than for an ELM without the pellet shown in figure 3b illustrating that the pellet triggered ELMs could be associated with larger particle loss compared to non-pellet ELMs. Note that the ELM affected area at the outer part of the plasma, $\delta r_{ELM}/a \sim 0.15$, is approximately the same as for the non-pellet ELM in figure 3b so that the difference is due to the ELM amplitude. As already mentioned a precise evaluation of particle loss is complicated due to the 3D character of the loss process. In this context it is interesting to note that the 3D perturbation (over-pressure "bump") caused by pellet has been observed during pellet-trigged ELMs on JET [14] and it is also reproduced in MHD modelling [15, 16].

Finally note that the size of the pellet deposition area is about $\Delta r_{pel}/a \sim 0.3$, which is similar to that expected in ITER [17, 18].

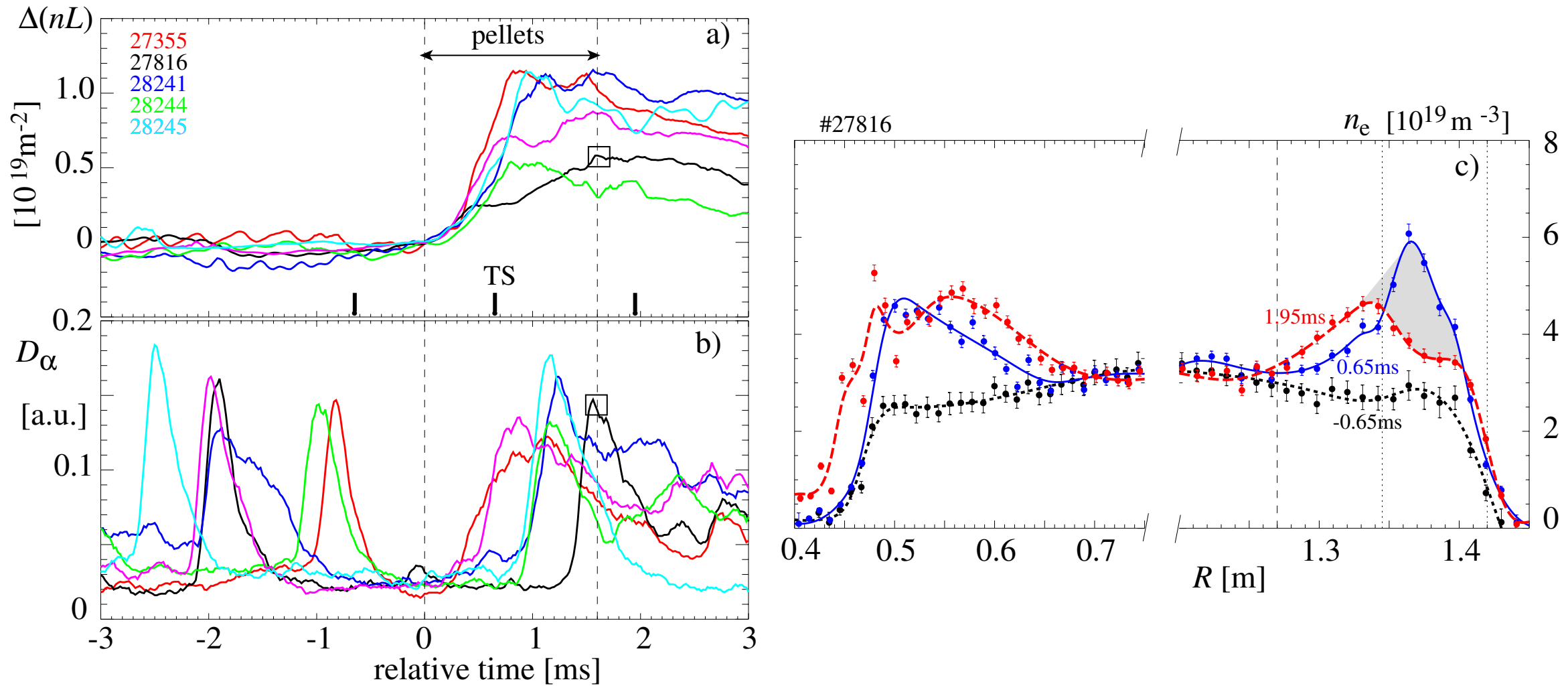

**Figure 4.** Synchronisation of pellets and ELMs. (a) increment of interferometer signal $\Delta(nL)$ during the pellets. (b) $D_\alpha$ emission from lower outer divertor target. (c) Density profiles for one shot marked by squares on $\Delta(nL)$ and $D_\alpha$ traces at times indicated by markers in panel (a).

## 6. Post-pellet particle losses

The second assumption of the idealised fuelling model shown in section 2 is that the post pellet losses occur due to ELMs of constant size and frequency which are the same as for mitigated ELMs without pellets.

Figure 5 shows the first example when these assumptions are not satisfied. It is seen that immediately after the pellet the density decays 4.8 times faster than the ideal rate calculated for a non-pellet case in eq (4). This fast particle loss is caused by a "compound" ELM, i.e. an ELM followed by a transient L-mode-like phase. The evolution of the density profile during this phase is shown in figure 5b. It is seen that the area affected by rapid particle loss encompasses the whole pellet deposition zone. It is also noticeable that the profile evolution comprises mainly outward loss and virtually no inward diffusion. The corresponding particle flux $\Gamma/n_e$ can be estimated from the continuity equation $\Gamma/n_e \sim -1/n_e \int_0^R \Delta n_e/\Delta t dR'$, where the time derivative is evaluated by $\Delta n_e/\Delta t$ from two subsequent density profiles during the density decay and the source term due to gas is omitted. The magnitude of this particle flux $\Gamma/n_e$ is shown in figure 5b. It is seen that the peculiar feature of the outward particle loss in the zone with positive density gradient ($R = 1.20 - 1.28$ m) can be explained by a convection with a velocity of $\sim 3$m/s. In the zone with conventional negative density gradient (say at $R = 1.34\text{m}$), the particle flux is $\Gamma / n_e \sim 10\text{m/s}$. If this flux is fully attributed to diffusion then the coefficient is $D \sim 1.5\ \text{m}^2/\text{s}$ ($L_n = 1/\partial_R \ln n_e \sim -0.15\text{m}$). The reason for this in–out asymmetry of post pellet transport is not known. This effect could be important for pellet fuelling of next step devices and therefore it needs to be understood.

For completeness figure 5a shows temporal evolution of the plasma energy $W_{mhd}$ determined by equilibrium reconstruction. At the onset of pellet-triggered "compound" ELM a small drop of $W_{mhd}$ can be observed but the main effect is the change in slope in temporal evolution of $W_{mhd}(t)$. Small change in $W_{mhd}$ during post-pellet density decay is consistent with the observable increase of electron temperature measured by Thomson scattering. In this context it is useful to note that initial density losses due to pellets (so called "first filament") can occur even without pellet triggered ELM [14, 15, 16].

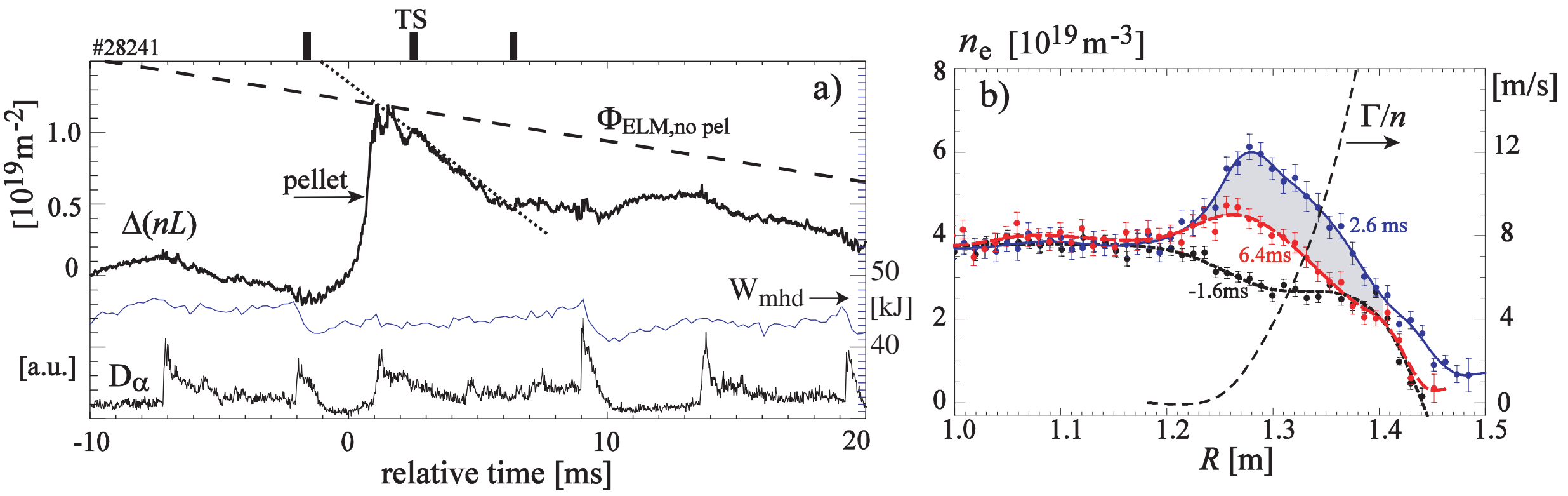

**Figure 5.** (a) Change in line integrated density $\Delta(nL)$, $D_\alpha$ emission and the plasma energy $W_{mhd}$ (blue solid line) for shot with pellet and mitigated ELMs. The density offset of $nL = 3.4 \times 10^{19}\ \text{m}^{-2}$ and the time offset of 0.3218s are subtracted. Broken line is the decay rate without pellet from equation (4) and dotted line is 4.8 larger than this rate. (b) Density profiles at times shown by markers on panel (a). Broken line is the normalised particle flux $\Gamma / n_e$. $I_p = 600\text{kA}$, $P_{NBI} = 3.6\text{MW}$.

The second example shown in figure 6 represents a favourable situation where the post pellet losses are similar to the ideal fuelling model. The size and frequency of the ELMs have not changed significantly (~20%) due to the pellet as seen from the traces of line integrated density, $D_\alpha$ emission and the plasma energy $W_{mhd}$. The averaged post-pellet density decay is approximately linear and is similar to that calculated from mitigated ELMs without pellets in equation (4). Density profiles are not available for this pellet, however, a bremsstrahlung image shows that the pellet evaporation zone is $r_{pel} \geq 0.85a$ (figure 6b). This is similar to pellet deposition in ITER and shows that the favourable post-pellet behaviour is not the result of deep pellet penetration. This is the closest scenario to that expected in ITER which has been obtained on MAST so far in terms of similarity of pellet/ELM ratio and post pellet density decay.

The reason for the difference in the examples shown above is not well understood. Both plasmas have identical RMP current and configuration, identical heating power, similar shapes and similar densities and pellet sizes. The most significant difference is that the favourable example has somewhat lower plasma current compared to the unfavourable case. It should be noted that in the unfavourable example (figure 5), "compound" ELMs are present also before the pellet. Nevertheless, during these ELMs, the density decay rate is not significantly higher than predicted by the ideal case as seen in figure 5a about 7 ms before the pellet.

In all analysis we have ignored the particle source from gas fuelling. The importance of this term in our plasmas is clearly seen from the spontaneous increase of plasma density during the inter ELM H-mode phases in examples shown in figures 3 and 5. An evaluation of the gas sources requires 2D simulations in order to account for the poloidal modulation of neutrals around the plasma. Such an analysis is outside the scope of the present paper and is planned in the future.

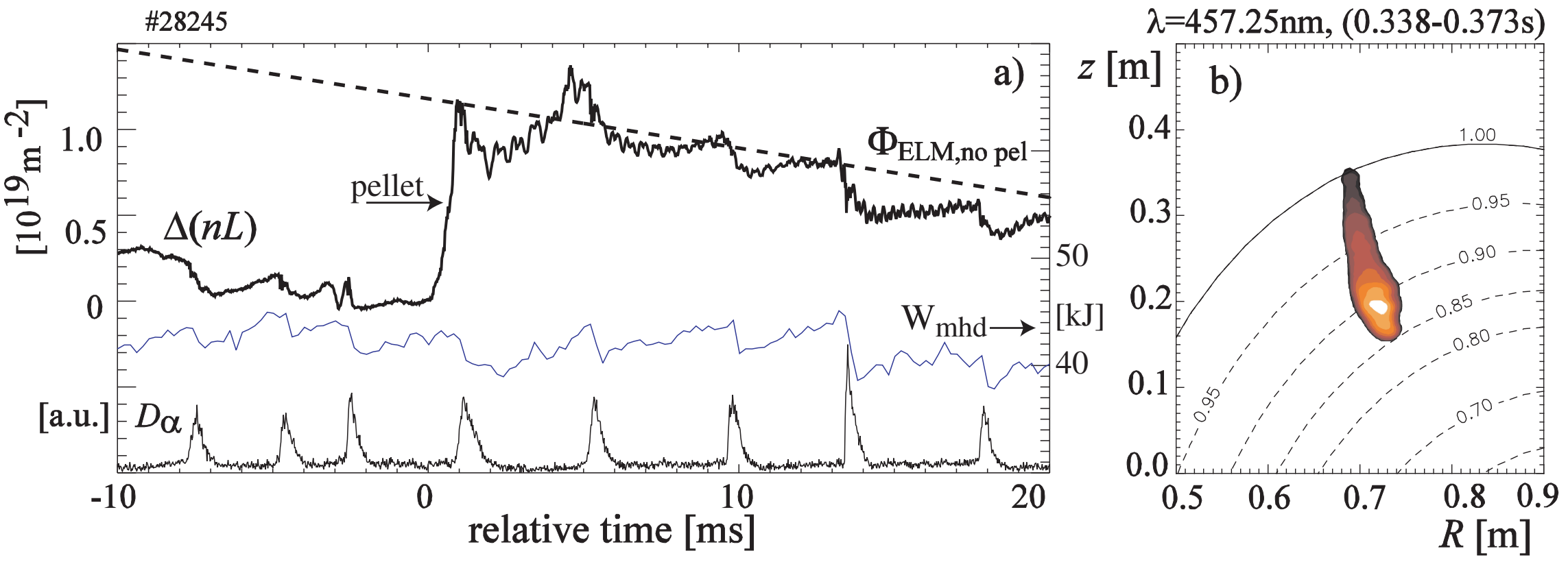

**Figure 6.** (a) Change in line integrated density $\Delta(nL)$, $D_\alpha$ emission and the plasma energy $W_{mhd}$ (blue solid line) for shot with pellet and mitigated ELMs. Density offset of $nL = 3.9\times10^{19}\,\mathrm{m}^{-2}$ and time offset of 0.3668s are subtracted. Broken line is the decay rate without pellet from equation (4). (b) open shutter visible bremsstrahlung image of the pellet. The contour labels are $r/a = \sqrt{\psi_N}$, where $\psi_N$ is the normalised poloidal magnetic flux. $I_p = 550\mathrm{kA}$, $P_{NBI} = 3.6\mathrm{MW}$.

## 7. Conclusions

This paper reports on the first experiments on MAST with simultaneous pellet fuelling and ELM mitigation by RMP coils. The data are compared with the pellet fuelling model which has been formulated for ITER. It is shown that the fuelling pellets trigger ELMs and their size could be larger than non-pellet induced RMP-mitigated ELMs. Concerning the post pellet loss an example is shown in which similarity with the ideal pellet fuelling model is demonstrated simultaneously in the following aspects:

- pellet deposition is ITER-like, $r_{pel}/a > 0.7$
- pellet to ELM particle ratio is ITER-like, $N_{pell}/\delta N_{ELM} \sim 6$
- post-pellet loss rate is constant and the same as due to mitigated ELMs without pellet

The relative ratio of pellet to plasma particle content is larger on MAST than on ITER as is the case in the majority of present devices.

Clearly a larger dataset is needed to understand particle losses under the condition of simultaneous pellet fuelling and ELM mitigation by RMPs. The aim is to demonstrate pellet fuelling with ELM mitigation which is simultaneously compatible with the plasma core, divertor and overall fuel balance. A high frequency pellet injector in MAST Upgrade would further increase the similarity with the ideal fuelling model, in particular by reducing the contribution of gas fuelling.

**Acknowledgement**

This work was funded by the RCUK Energy Programme under grant EP/I501045 and the European Communities under the contract of Association between EURATOM and CCFE. The views and opinions expressed herein do not necessarily reflect those of the European Commission. Authors would like to thank Dr B. Lloyd for valuable comments. Anonymous referees are gratefully acknowledged for their suggestions.